# Accelerated Atomistic Simulations of a Supramolecular Polymer in Water

*Davide Bochicchio and Giovanni M. Pavan**

Department of Innovative Technologies, University of Applied Sciences and Arts of Southern Switzerland, Galleria 2, CH-6928 Manno, Switzerland

giovanni.pavan@supsi.ch

**ABSTRACT:** All-atom molecular dynamics has been recently proven a useful tool for the study of supramolecular polymers. While the high resolution offered by the atomistic models may allow for deep comprehension of the assembled structure, obtaining a reliable equilibrated configuration for these soft assemblies in aqueous solution remains a challenging task mainly due to the complexity of the atomistic systems and the long simulation time needed for their equilibration. Here we have tested two well-known advanced simulation methods (Accelerated Molecular Dynamics and Metadynamics) on an atomistic model of water-soluble 1,3,5-benzenetricarboxamide (BTA) supramolecular polymer and have compared the obtained data to classical molecular dynamics. Our results show that both techniques can be very useful to accelerate the equilibration of the supramolecular fiber in solution. Moreover, we demonstrate that combining the two methods in opportune way allows to take advantage of the strong points of both, providing an equilibrated configuration for the BTA supramolecular polymer in a faster and more reproducible way compared to classical simulations. The versatility of this approach suggests that the latter can be adapted to simulate a variety of supramolecular polymers as well as different types of supramolecular assemblies.

**INTRODUCTION**

Supramolecular polymers, where the monomers are interconnected via non-covalent interactions, have recently received notable attention thanks to their dynamic and adaptive properties.[1–3] These self-assembled structures possess self-healing behavior, stimuli responsiveness and dynamic bio-inspired properties that are very promising for the creation of novel advanced materials for bio- and nano-applications.[4] In this context, gaining deep understanding of the non-covalent interactions between the monomers is a first fundamental step toward the creation of supramolecular polymers with controlled properties.[5,6] However, the experimental characterization of these soft assemblies is extremely difficult, especially in water, because of their reduced size and the limited contrast offered in solution. This produces a general lack of molecular-level insight into their structure, while in most cases it is almost impossible to obtain from the experiments information on the complex interplay between the different types of interactions (H-bonding, electrostatic, van der Waals, hydrophobic, etc.) controlling the assembly in the aqueous environment.

Given these clear limitations, molecular simulations have recently become an important tool to investigate the structure and properties of supramolecular polymers.[6–22] An extensively studied case is constituted by 1,3,5-benzenetricarboxamide (BTA) supramolecular polymers. The BTA monomers directionally self-assemble into supramolecular fibers due to three-fold hydrogen bonding between the amides and stacking of cores(see Figure 1a-c).[23] Previous computational studies based on density functional theory (DFT) focused on the cooperativity of inter-monomer H-bonding in small BTA stacks in the gas-phase.[9,10,19] H-bonding and dipole amplifications into stacks of BTA monomers with short side chains have also been studied by all-atom classical Molecular Dynamics (MD) simulations in organic solvents.[11] A coarse-grained model of such reduced size BTA derivatives has also been recently reported and tested, which allowed to study BTA self-assembly in nonane.[20] While all these computational efforts provided important insights into the BTA assembly in intrinsically ordered conditions (*i.e.*, the apolar solvent and the semi-rigid nature of the monomers favor ordered stacking), complexity increases enormously when moving to water, mainly due to the more complex and dynamic structure of the water-soluble monomers (Figure 1a,b) and to the strong hydrophobic effects that compete with intra-monomer H-bonding.[21]

In 2011 the group of Schatz studied a supramolecular polymer composed of peptide amphiphilic monomers in explicit water by means of classical MD simulations.[7] Their MD results provided a detailed picture of the structure of the polymer and H-bonding between the monomers in the assembly. Recently, our group used all-atom MD simulations to study a BTA supramolecular polymer in explicit water (Figure 1). This model of an infinite (achiral) BTA fiber was compared to a second polymer variant, where the addition of a single methyl group into the side chains of the BTAs (making the monomers chiral) imparted preferential helicity to the supramolecular fiber.[8] These atomistic models provided important insight into the assembly structure, stability, monomer-monomer interactions, competition of hydrophobic effects *vs.* H-bonding and penetration of water into the fiber structure.[8] The MD results were also found in optimal agreement with the experiments, providing evidence for stable H-bonding between the BTA cores and structural and energetic data useful to rationalize experimentally observed differences in the structure, stability and dynamics of the achiral and chiral BTA supramolecular polymers.

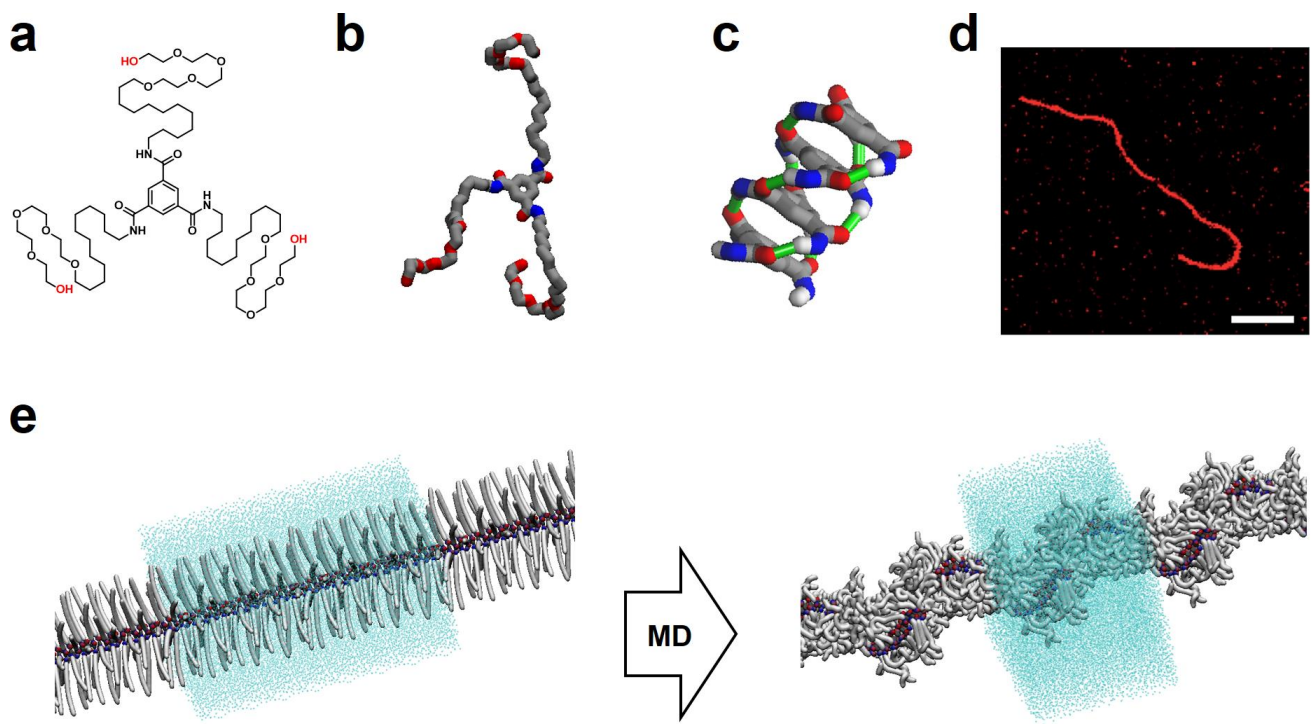

**Figure 1.** Water-soluble BTA supramolecular polymer. (a) Chemical structure of a water-soluble BTA achiral monomer[8,24]. (b) Atomistic model for the BTA monomer. Color code: carbons in gray, oxygens in red and nitrogens in blue. Hydrogens are not shown for clarity. (c) Detail of four stacked BTA cores showing the threefold H-bonding (in green) between the cores. (d) STORM image of the supramolecular fibers in water (scale bar, 2 μm).[8,25] (e) Folding of an infinite BTA supramolecular fiber model in water from all-atom MD simulation (**MD$_0$**)[8].

In general, due to the lack of a precise experimental structure for these complex statistical materials, the first fundamental simulation challenge is to obtain from the models a reliable structure for the supramolecular polymer in water and in equilibrium conditions. Typically, atomistic MD simulations start from an initial reasonable configuration for the supramolecular assembly[7,8] created based on the available information from experiments or from higher levels of theory (*i.e.*, DFT).[9] The starting configuration is then relaxed by means of MD to obtain an equilibrated structure for the supramolecular polymer.[7,8,26] The latter is a non-trivial task, given the large size and the complexity of these atomistic systems. Although simulation times of hundreds of nanoseconds are nowadays accessible also for the all-atom MD simulation of such large systems (size also exceeding a few 100k atoms), critical points remain (i) the slow dynamics of the supramolecular structure to reach an equilibrated configuration in the solvent and (ii) the intrinsic sampling limitations of classical MD in studying such complex structures, where during the MD run the assembly can remain trapped in local energy minima. For these reasons, reaching the "true equilibrium" configuration for such complex systems by means of all atom models and classical MD still represents a rather challenging goal.

A possible approach to overcome the aforementioned limitations is to coarse-grain the system, simplifying the structural description of the monomers to be able to reach longer time scales. Recently, coarse graining has been successfully used to study a variety of supramolecular systems.[13,16–18,20]

In this work we investigate a different approach: the use of advanced sampling techniques to accelerate the equilibration of complex all atom supramolecular systems. Two popular advanced simulation methods, Accelerated Molecular Dynamics (aMD)[27] and Metadynamics (MetaD)[28,29], have been tested on an atomistic model of water-soluble BTA supramolecular polymer (Figure 1) that was previously investigated by means of classical MD.[8] Accelerated Molecular Dynamics (aMD), reported by McCammon in 2004,[27] has proven an effective method to enhance the conformational sampling of proteins and dendrimers in water.[30–34] MetaD has previously proven useful for a vast variety of systems, including for instance the study of complex synthetic branched macromolecules in solution (*i.e.*, dendrimers).[35,36] Importantly, our data demonstrate that these approaches can be useful to (i) accelerate the equilibration dynamics of a supramolecular system and to (ii) increase the confidence in the reliability of the equilibrium configuration obtained with these advanced simulation techniques.

**RESULTS AND DISCUSSION**

Our previous all-atom MD simulations of the BTA supramolecular polymer shown in Figure 1e started from an initial model for the BTA fiber consisting of 48 initially extended BTA monomers stacked into a simulation box filled with explicit TIP3P water molecules.[37] This system forming an infinite supramolecular polymer along the main fiber axis through periodic boundary conditions (Figure 1e, left) was relaxed in NPT conditions using anisotropic pressure scaling to allow free rearrangement of the supramolecular fiber during the equilibration MD.[8] Herein, we will identify this reference classical MD simulation as **MD$_0$**. Early during **MD$_0$** the side chains of the BTAs collapsed around the hydrophobic BTA cores to minimize the contacts with water (primary folding), while the whole initially extended fiber started to fold (secondary folding) to further reduce the hydrophobic surface exposed to the solvent. In this case, 400 ns of **MD$_0$** were needed to reach a configuration (Figure 1e, right) in which several key structural and energetic parameters used to characterize the fiber (fiber length, monomer-monomer interaction energy, radial distribution functions $g(r)$ of the cores, H-bonds, etc.) showed good convergence in the MD regime.[8] Theoretical small-angle X-ray scattering (SAXS) profiles obtained from the equilibrated phase MD simulation (the last 100ns of **MD$_0$**) were also found in good agreement with the experimental ones.[8] However, while apparently fiber folding came to an end and the fiber appeared as fully equilibrated during **MD$_0$**, the structure still presented hydrophobic patches exposed to the solvent and a rather wrinkled and irregular surface. The average fiber cross-sectional radius obtained from **MD$_0$** was also found slightly smaller than the SAXS experimental one (3.1±0.2 nm). Considering the limited sampling efficiency of classical MD, especially for very complex systems, these observations suggested that fiber folding could have possibly been still incomplete.

**Accelerated MD simulations of a BTA supramolecular polymer in water.** The fundamental idea behind the aMD technique is to raise the potential wells of the system to allow faster crossing of the energy barriers dividing the different configurations that the system can access.[27] This is done by adding an energetic bias to the global potential of the whole system, to the sole dihedrals bonded terms or to both (global potential plus an extra bias term dedicated to dihedrals) when these are below certain threshold values.

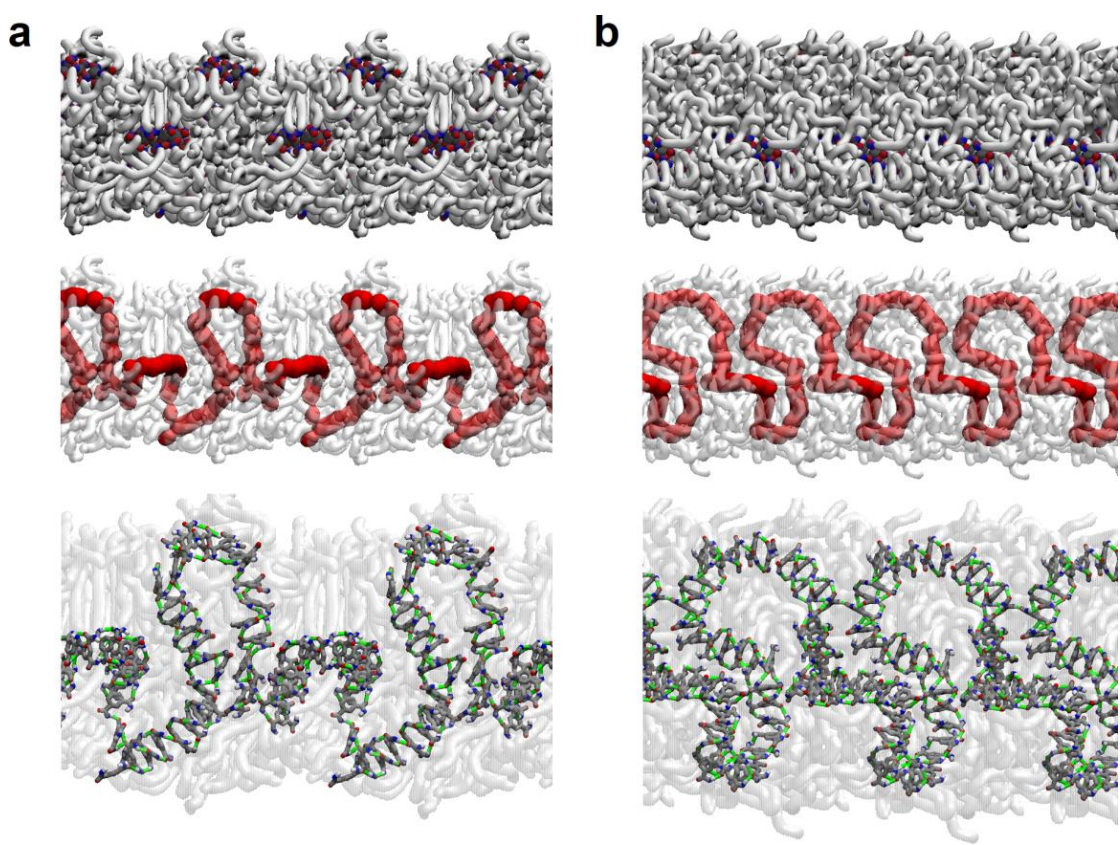

**Figure 2.** Equilibrated configurations for the BTA fiber at different levels of detail obtained from **aMD$_0$** (a) and **MetaD-aMD$_{MD}$** (b) simulations. Top: view of the full fiber (same color scheme as Figure 1e). Middle: The stacked cores are represented as a continuous red surface, while the rest of the fiber is transparent. Bottom: zoom into the fiber showing the H-bonding network (green), with carbon, oxygen, nitrogen and hydrogen atoms of the cores respectively colored in gray, red, blue and white. Explicit water molecules are not shown for clarity.

Here we have used the aMD method as implemented in AMBER 14[38] to simulate the BTA supramolecular polymer of Figure 1e and we have compared the results with those obtained from classical **MD$_0$**. We have applied the bias to both the whole potential and the dihedrals, where the bias parameters have been calculated using the standard approach used in previous aMD studies reported in literature (see Materials and Methods).[27,30–33] First we ran an aMD simulation starting from the same initially extended configuration of the BTA fiber used as a starting point for **MD$_0$** (Figure 1e, left). We will identify this aMD simulation **aMD$_0$**.

During **aMD$_0$** the BTA fiber undergoes similar primary and secondary folding than in **MD$_0$**, but the process was sensibly faster. Interestingly, fiber folding did not stop at the **MD$_0$** level, but proceeded further for ≈200 ns of **aMD$_0$** before coming to an end. The **aMD$_0$** run was then prolonged for additional 200 ns to ensure full equilibration in the aMD regime, during which we did not observe any relevant structural or energetic change. The final structure of the fiber produced by **aMD$_0$** is shown in Figure 2a. Compared to **MD$_0$**, **aMD$_0$** produced a more compact/uniform and nearly cylindrical fiber while substantially preserving the order of the BTA cores. In fact, the latters are found as stably stacked into a rather continuous serpentine with a few punctual defects emerging (Figure 2a).

We investigated to what extent the final equilibrated state for the BTA fiber produced by **aMD$_0$** was different from that obtained from **MD$_0$**. In particular, we focused on structural features as well as on the main factors controlling the BTA assembly in water (hydrophobic effects, H-bonding, etc.).[8] A good indicator of the hydrophobic effects involved in fiber folding can be obtained from the solvent accessible surface area (SASA) of the BTA supramolecular polymer – *i.e.*, the stronger the SASA shrinkage at the equilibrium (after fiber folding), the better the fiber rearranged its structure to minimize the contacts with water. To elucidate the differences between **aMD$_0$** and **MD$_0$**, we plotted the two simulation trajectories in a 2D space (Figure 3a: **MD$_0$**, black; **aMD$_0$**, cyan), where the *x* and *y* axes respectively report the average radius and the SASA of the BTA fiber during the runs. From the 2D plots of Figure 3a it is clear that in both simulations fiber folding is accompanied by strong SASA reduction, which highlights the dominant character of the hydrophobic effects. Accordingly, augmented folding produces an increase in the average fiber radius. In both cases the systems fall in a minimum where fiber's SASA and radius appear as equilibrated and oscillate around their average values. However, the two $MD_0$ and $aMD_0$ minima are significantly separated in the radius-SASA space – respectively, the equilibrated fiber configuration produced by $MD_0$ lies between ≈220-260 $nm^2$ of SASA and ≈2.1-2.3 nm of radius, while that from $aMD_0$ between ≈160-190 $nm^2$ of SASA and radius ≈2.9-3.1 nm. The stronger SASA shrinkage seen in $aMD_0$ suggests that the BTA fiber was able to better rearrange its structure to further reduce the contacts with the solvent compared to $MD_0$. The final fiber radius obtained from $aMD_0$ was also found to fit very well with the SAXS experiments (3.1±0.2 nm, in grey in Figure 3a),[8] whereas that obtained from $MD_0$ was found smaller – a point in favor of aMD.

In principle, proper reweighting[39] of the equilibrated phase $aMD_0$ trajectory would allow to estimate the free-energy surface (FES) of the BTA fiber in the radius-SASA space. While it is known that obtaining precise reweighting of the aMD trajectories can be problematic for such complex systems,[40] it is worth to stress that here we are not interested in obtaining a rigorous estimation of the FES (which, by the way, seems to possess a unique global minimum in the radius-SASA space, as confirmed by the constant average SASA and radius in the last part of $aMD_0$ and by the other simulations performed – see later on), but just to speed-up the equilibration of the system. While the aMD method has been here used with this sole purpose, we delegated more quantitative energetic and structural analyses on the equilibrated BTA fibers to additional 100 ns of unbiased standard MD simulation ($MD_{post}$) starting from the end of each biased simulation.

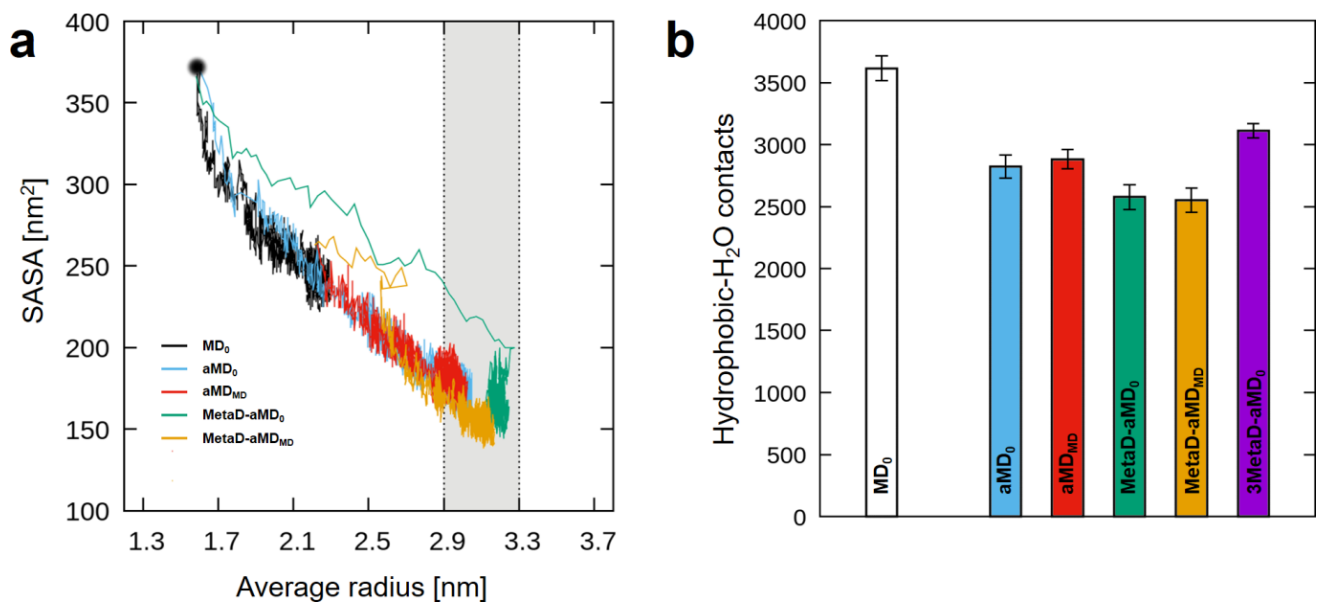

**Figure 3.** Results of the accelerated simulations of the BTA fiber in water. (a) Simulations trajectories in the radius-SASA space. The SAXS experimental range for fiber radius (3.1±0.2 nm) is colored in gray. (b) Average number of contacts between the hydrophobic parts of the fiber (cores and dodecyl spacers) and water molecules.

During production $MD_{post}$ the fiber did not change its structure anymore, indicating good stability for the final configuration produced by **aMD₀**. The same approach has been followed also for all other enhanced sampling simulations tested in this work (see below). We calculated the number of contacts between water and the hydrophobic regions of the fiber (BTA core and dodecyl spacers). Coherent with the SASA data, these are found reduced in **aMD₀** compared to **MD₀** (Figure 3b). Another parameter that goes in the same direction is the radial distribution function $g(r)$ of the water molecules calculated respect to the BTA cores (Figure 4a). This data represents the probability of finding water molecules within a certain distance from the cores. We observe a reduction of ≈50% of water penetration within 1 nm from the BTA cores in **aMD₀** compared to **MD₀**. All these data suggest that a key driving force for the augmented folding seen in **aMD₀** (and also in all other biased simulations, see below) lies in the hydrophobic effects.

Another important analysis concerns the level of order in the BTA core stacking, which is a good descriptor of the stability and persistency of the supramolecular assembly.[8,21,26] This can be monitored by calculating the radial distribution function, *g(r)*, of the BTA cores as a function of the inter-core distance, which peaks (at distances multiple of the stacking one, ≈0.35 nm) are the fingerprints of core stacking. Compared in Figure 4b, the core-core *g(r)* plots are found similar in **MD$_0$** (black) and **aMD$_0$** (cyan), indicating that the core stacking order has been substantially preserved in **aMD$_0$**. The first *g(r)* peak is however slightly reduced, consistent with the presence of some "broken" points emerging along the fiber early during the **aMD$_0$** folding.

We have compared the energies of the equilibrated BTA fiber as obtained from **MD$_0$** and **aMD$_0$**. For what concerns the total potential energy of the fiber, only small differences could be detected. Consistent with what said above for the core-core *g(r)* plots, this indicates that the fiber as seen during the last 100 ns of **MD$_0$** was already almost equilibrated from the point of view of the monomer-monomer interactions and overall persistency/stability of the BTA assembly. In Table 1 we report the main features of the BTA supramolecular fiber that are found improved in **aMD$_0$** compared to **MD$_0$**. For example, as anticipated, the equilibrated fiber SASA is found reduced in **aMD$_0$** compared to **MD$_0$** by ≈27% – ΔSASA = SASA(**aMD$_0$**) – SASA(**MD$_0$**) in Table 1 –, which confirms that the fiber can undergo better compaction during **aMD$_0$** compared to **MD$_0$** to optimize its hydrophobic/hydrophilic interactions. The electrostatic term of BTA-BTA interaction (Δ$E_{ele}$) is also found more favorable in **aMD$_0$** than in **MD$_0$** by ≈28% – Table 1: ΔΔ$E_{ele}$, calculated as Δ$E_{ele}$(**aMD$_0$**) – Δ$E_{ele}$(**MD$_0$**), equal to ≈–5.8 kcal/mol per-BTA monomer (see also SI). This is coherent with an overall stabilization of the H-bonding network in the fiber, where the average number of H-bonds per-BTA is seen to increase by ≈20% in the case of **aMD$_0$**. These observations suggest that under different aspects the equilibrated state reached by the BTA fiber during **aMD$_0$** is energetically more optimized compared to that from classical **MD$_0$**.

**Table 1.** Main improved features in the equilibrated BTA fiber configurations provided by the biased simulations compared to classical **MD₀**. Per-monomer $\Delta\Delta E_{ele}$ values are expressed in kcal/mol. $\Delta$SASA values are expressed in nm$^2$. All data have been collected from 100 ns of unbiased MD simulation (MD$_{post}$) following to each biased simulation and compared ($\Delta$) with the last equilibrated phase (100 ns) of classical **MD₀**.

| Comparison with **MD₀** | **aMD₀** | **aMD$_{MD}$** | **MetaD+aMD₀** | **MetaD+aMD$_{MD}$** | **3MetaD+aMD₀** |
|---|---|---|---|---|---|
| $\Delta\Delta E_{ele}$[a] (kcal/mol) | -5.79 (+28%) | -5.52 (+27%) | -4.17 (+20%) | -7.53 (+36%) | -7.55 (+37%) |
| $\Delta$H-bonds[b] per-BTA (#) | +0.30 (+20%) | +0.48 (+32%) | +0.08 (+5%) | +0.34 (+22%) | +0.31 (+21%) |
| $\Delta$SASA[c] (nm$^2$) | -67 (+27%) | -63 (+26%) | -87 (+36%) | -91 (+37%) | -95 (+39%) |

[a]$\Delta\Delta E_{ele}$ is the difference between the electrostatic component of the BTA self-assembly energy ($\Delta E_{ele}$) calculated from MD$_{post}$ (following to each biased simulation) and from **MD₀** (values per-BTA monomer). [b]$\Delta$H-bonds parameter is the difference between the average number of H-bonds per-BTA calculated from MD$_{post}$ and from **MD₀**. [c]$\Delta$SASA is the difference between the SASA of the fiber calculated from MD$_{post}$ and from **MD₀**.

To increase our confidence in the aMD approach, we performed a second aMD simulation (**aMD$_{MD}$**) using the same bias as in **aMD₀** but starting from a different configuration for the fiber: the pre-equilibrated one obtained from **MD₀** (Figure 1e, right). In this way we intended to verify if this method was able to produce a similar equilibrated state for the fiber while starting from two different configurations.

In **aMD$_{MD}$** we observed once again that the BTA fiber folded further than in **MD₀**. In this case fiber folding reached completion after ≈150 ns of aMD, while the **aMD$_{MD}$** run was prolonged for other 250 ns to verify the convergence of the structure. At the equilibrium, **aMD$_{MD}$** provided a final radius and an equilibrated SASA for the fiber substantially identical to those obtained from **aMD₀**, as demonstrated by the superimposition of the red and cyan minima at the end of the trajectories reported in Figure 3a. Again, we run additional 100 ns of classical MD$_{post}$ after **aMD$_{MD}$** to perform all analyses conducted for the other cases (data in Figures 3, 4 and in Table 1). Aside from subtle differences imputable to normal statistical oscillations, all the extracted parameters indicate that **aMD$_0$** and **aMD$_{MD}$** produced almost equivalent final states for the fiber. This is demonstrated by the core-core *g(r)* plots (Figure 4a) showing again nearly identical core stacking in **aMD$_{MD}$** (red) and in **MD$_0$** (black). Moreover, all data from **aMD$_{MD}$** pertaining to fiber hydration are substantially identical to those of **aMD$_0$**. Similar to **aMD$_0$**, **aMD$_{MD}$** also shows analogous improvements compared to **MD$_0$** for what pertains to the average number of H-bonds per-BTA and monomer-monomer electrostatic interaction (Table 1), demonstrating the reliability of the aMD method in producing results that are substantially independent from the starting configuration.

**Metadynamics.** We decided to test also a different biased method, Metadynamics,[28,29] for two main purposes: (i) to check if a different biased approach could produce similar results and (ii) to try to accelerate further the fiber folding process. In fact, the latter took considerable simulation time to reach the equilibrium (>200 ns) even in the aMD accelerated regime. Metadynamics (MetaD) is a very popular biased approach which main purpose is to allow crossing the free-energy barriers between different states and to enhance the sampling of the conformational space of the system. For instance, MetaD has been recently proven useful to allow a more exhaustive sampling compared to classical MD of complex hyperbranched polymers (dendrimers) in aqueous solution.[35,36] The MetaD method requires the identification of one (or more) collective variable (CV) able to describe the slow degree of freedom of the system that has to be accelerated. However, the identification of the right CVs is known to be a key and often non-trivial point.[29]

While MetaD is typically used to obtain an accurate description of the FES of the simulated system, we stress again that here our main scope in using MetaD was not to obtain a precise sampling of the FES of the BTA fiber, but rather to speed-up in reliable way the slowest macroscopic process (fiber folding) leading to the equilibration of the supramolecular BTA fiber in the solvent. After a few tests, we identified the best suited CV to our aim: the maximum *z* distance between the centers of mass of the BTA cores, strictly connected to the fiber elongation/folding along its main axis (details on the MetaD parameters used in this work can be found in the Materials and Methods section).

The results of our first tests with MetaD were surprisingly good for what pertained to the folding speed. Starting from the initially extended configuration for the BA fiber (Figure 1e, left), the folding was strongly accelerated compared to aMD and the fiber *z* dimension (fiber length) went straight down to values similar to the equilibrium ones of the aMD runs, but in well reduced time. Only ≈5 nanoseconds of MetaD were sufficient to produce comparable fiber folding and average cross-sectional radius to those obtained from **aMD$_0$** and **aMD$_{MD}$** in hundreds of nanoseconds (150 ns and 200 ns respectively).

Despite the notable folding acceleration produced by MetaD, the core-core stacking (*g(r)*) was not destabilized, but on the contrary it was found better preserved than in the aMD simulations and nearly identical to **MD$_0$**. This indicated that the MetaD method allowed to fold the fiber in faster but at the same time more delicate way than aMD, at least in the earlier stages of fiber folding (*i.e.*, very far from the equilibrium). However, one disadvantage of this MetaD approach was that fiber folding was so fast that no sufficient time was left for full reorganization of the BTA lateral chains. From this viewpoint, the final configuration is still reminiscent of the starting one, and this actualized in a non-uniform rough surface and sub-optimal SASA that turned out very difficult to be optimized in efficient way at such accelerated regime and using this sole CV.

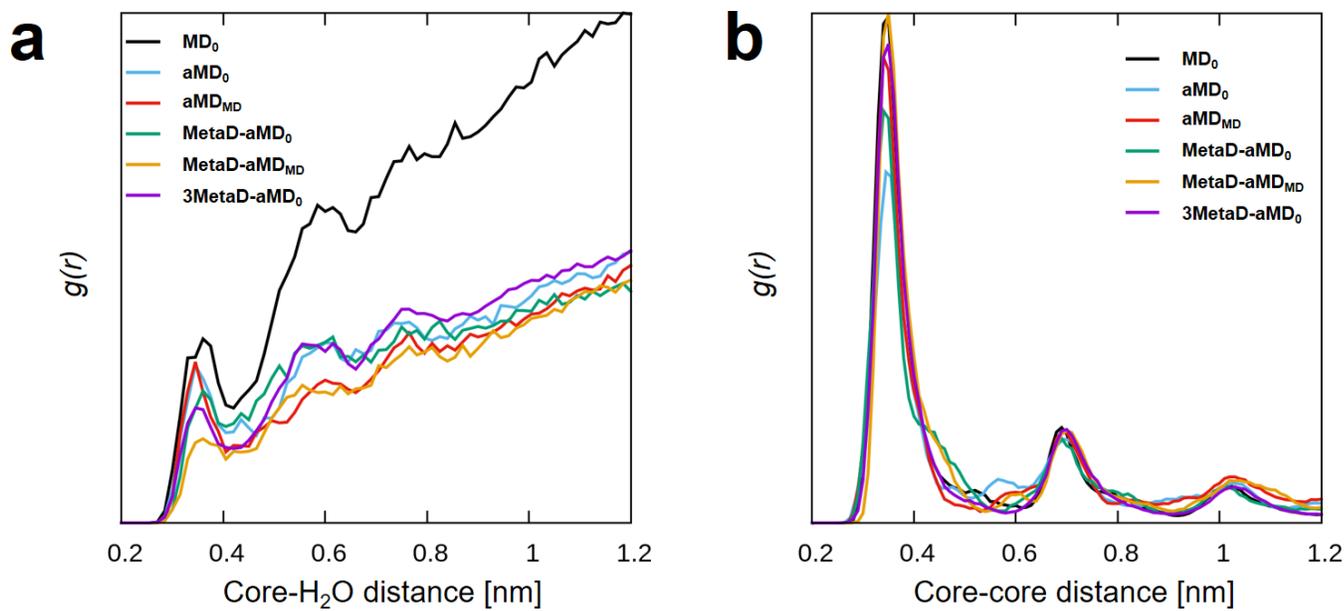

**Figure 4.** Radial distribution functions, *g(r)*, of (a) the water molecules calculated respect to the BTA cores (water penetration into the fiber) and of (b) the BTA cores as a function of the core-core distance.

**A combinined MetaD-aMD approach**. In light of these observations, we decided to take advantage of the strong points of both methods using them in a combined protocol. Namely, (i) we first used MetaD to accelerate the "gross" folding of the initially extended fiber and then (ii) we used aMD to promote fiber rearrangement and to optimize the folded state. As previously done for aMD, we conducted two simulations (**MetaD-aMD$_0$** and **MetaD-aMD$_{MD}$**) starting from two different configurations for the fiber: the initially extended BTA fiber and the pre-equilibrated configuration from **MD$_0$** reported in Figure 1,left and Figure 1,right respectively. As a criterion to end the first phase (MetaD) and chose a starting point for the second one (aMD) we used a very simple one: we took the configuration of the system immediately after the abrupt change in the slope of the MetaD hills seen during the run (see SI) – namely, the point where the fiber reaches the maximum folding level along the *z* axis. Obtaining these folded yet non-optimized fibers required just ≈5-7 ns of MetaD simulation. Then, these pre-folded configurations underwent 150 ns of aMD, sufficient to optimize the BTA side chains reaching full equilibration of the BTA fibers from the structural and energetic points of view (see SI).

The trajectories obtained from the combined **MetaD-aMD$_0$** and **MetaD-aMD$_{MD}$** simulations are also reported in Figure 3a (green and yellow respectively). Qualitatively, these show that the fibers follow slightly different paths during these equilibration runs compared to pure aMD simulations, first increasing the fiber average radius (MetaD: fast folding), and then optimizing their SASA (aMD: overall rearrangement). Seen in Figure 3a, the final fiber radii obtained from **MetaD-aMD$_0$** and **MetaD-aMD$_{MD}$** were found well compatible with the experimental one (grey region) and with the **aMD$_0$** and **aMD$_{MD}$** results. The final structure of the fiber was very similar too (Figure 2b: **MetaD-aMD$_{MD}$**), and detailed analyses of these simulations showed similar improvements compared to **MD$_0$** than in **aMD$_0$** and **aMD$_{MD}$** (Table 1). The ΔSASA (hydrophobic effects) and $\Delta\Delta E_{ele}$ (BTA-BTA electrostatic interactions) were found also slightly improved compared to the pure aMD simulations (by an additional ≈10%). Also, consistently with what said above, the first peak in the *g(r)* of the BTA cores as calculated from **MetaD-aMD$_0$** and **MetaD-aMD$_{MD}$** was found in general slightly higher than the corresponding **aMD$_0$** and **aMD$_{MD}$** simulations (Figure 4b). Considering statistical fluctuations, we should carefully say that the two approaches produce comparable equilibrated configurations for the BTA fiber. However, it is worth noting that the combined MetaD-aMD approach allowed to reduce the overall simulation time necessary to obtain an equilibrated configuration for the fiber by ≈75% compared to pure aMD. Thus, summarizing, every procedure used to accelerate the equilibration of the BTA supramolecular structure produced similar final configurations in terms of energy, SASA, and fiber radius (see Figure 3), significantly different from the final configuration obtained with classical MD. This means that our accelerated simulations demonstrated the ability to overcome local minima that are problematic for classical MD, augmenting the reliability of the final equilibrated configuration.

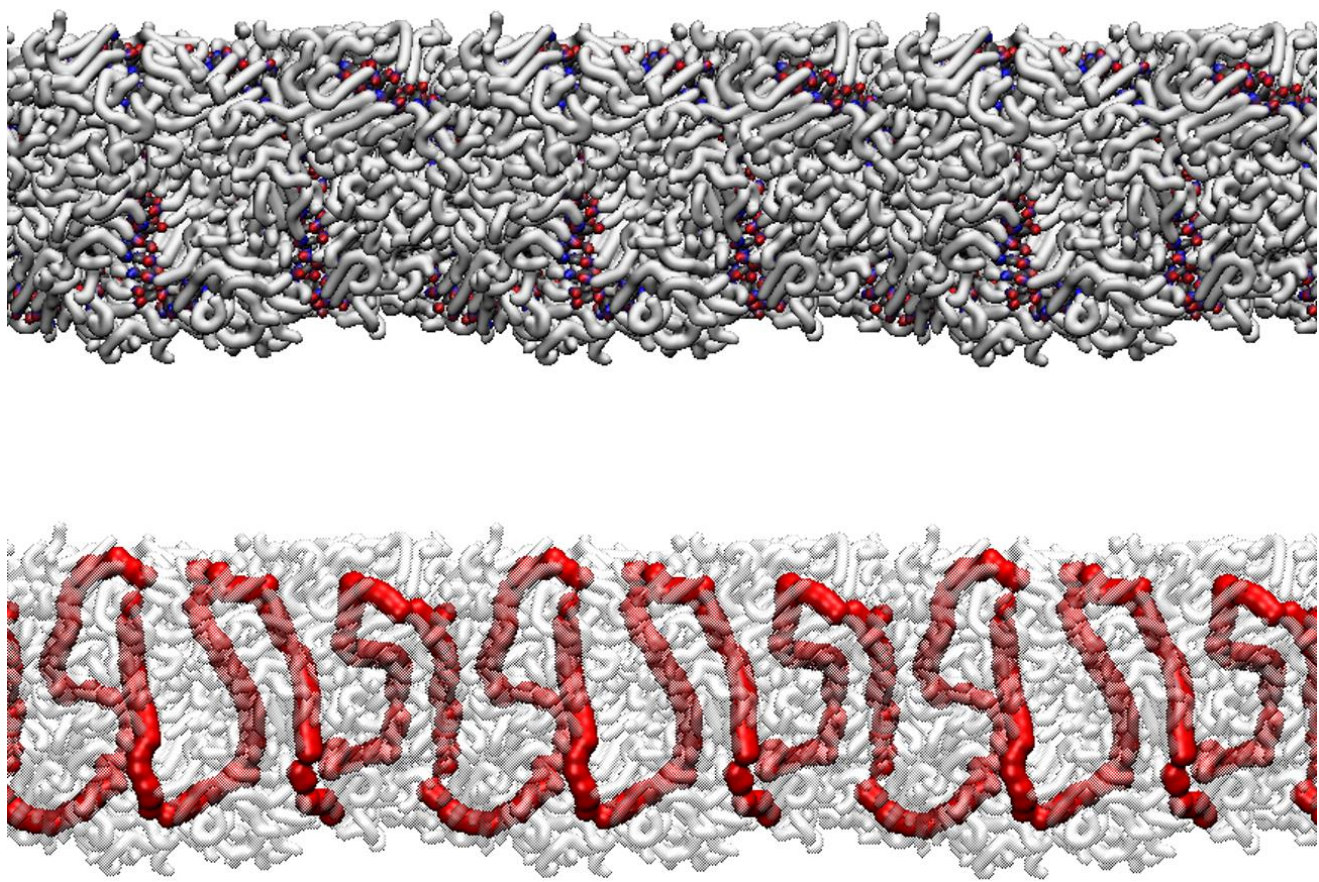

**Figure 5.** Equilibrated configuration of a BTA fiber (144 BTA monomers) obtained from the **3MetaD-aMD$_0$** simulation. The color schemes are the same as Figure 2.

Thanks to the technical advantages offered by the combined MetaD-aMD approach, we could also create and simulate (**3MetaD-aMD$_0$**) a BTA fiber three-times longer (144 BTA cores) than that simulated before.[8] This simulation was useful to prove that the overall characteristics of the BTA supramolecular polymer extracted from our 48 BTA polymer model were not influenced to some extent by examining a rather reduced section of the fiber (or, in other words, to exclude spurious PBC effects). This model was built starting from MetaD folding of the initially extended 48 BTA fiber, which was then triplicated (144 BTA) along the *z* axis and underwent 150 ns of equilibration aMD. The final equilibrated configuration for this larger system is shown in Figure 5. As previously done for all other systems, 100 ns of unbiased MD$_{post}$ simulation were run for data analyses.

All structural and energetic parameters demonstrate perfect agreement with the smaller 48 BTA systems (see Table 1 and Figure 4), proving that the combined MetaD-aMD approach can be remarkably useful (i) to obtain a reliable and well-reproducible equilibrium configuration for the BTA fiber in water, (ii) to dramatically speed-up fiber folding avoiding entrapment in local energy minima and as a consequence (iii) to simulate larger systems allowing for a more accurate statistical modeling of these supramolecular polymers.

**CONCLUSIONS**

In this work we have used two advanced simulation approaches, aMD and MetaD, to obtain the equilibrium configuration for an atomistic model of BTA supramolecular polymer in explicit water in efficient and reliable way. Our data highlight the technical advantages offered by both methods in the simulation of a very complex supramolecular system such as the water-soluble BTA fibers studied herein. In particular, our results demonstrate that a combined MetaD-aMD approach, where MetaD is first used to accelerate the initial fiber folding and aMD is then used to promote optimization of the supramolecular structure, allows for remarkable speed-up of the fiber equilibration process. Comparison of our simulations starting from different configurations for the BTA fiber and using two different biased techniques proves that this approach can produce an equilibrated configuration of the supramolecular BTA polymer in water that is reliable and well reproducible. This offers notable technical advantages respect to classical MD especially in terms of efficiency (reduced simulation time), possibility of studying larger systems and confidence in providing a high-resolution picture for the structure of these complex supramolecular systems in relevant environments and in equilibrium conditions. Our results also suggest that this simulation approaches can be opportunely adapted for the simulation of soft supramolecular assemblies other than supramolecular polymers, such as, for example, small micelles,[41,42] macrocycles and nanocages[43] to name a few.

**MATERIALS AND METHODS**

The entire simulation work was conducted using the AMBER14 software (classical MD and aMD) and the Gromacs 5.1.2 suite[44,45] patched with the PLUMED2[46] plugin for Metadynamics (MetaD) simulations. The molecular models for the BTA monomer and for the infinite BTA polymer simulated herein were taken from our previous MD work.[8]

All simulations have been performed using a time step of 2 fs and an 8 Å cutoff. Simulations in NPT conditions were run at room temperature (T=20 °C) and 1 atm of pressure. Anisotropic pressure scaling has been applied through the Berendsen barostat[47] with a time constant of 2 picoseconds. Langevin[48] and V-Rescale[49] thermostats have been used respectively within AMBER and Gromacs, in both cases with a time constant of 2 picoseconds. The SHAKE[50] and LINCS[51] algorithms have been used in AMBER and Gromacs respectively to constrain bonds involving hydrogens, while in all cases the Particle Mesh Ewald (PME) was used to treat long range electrostatics.[52]

The bias parameters for the whole potential energy (*AlphaP*) and for the dihedral term (*AlphaD*) used in all our aMD simulations have been calculated with the standard empirical formulas previously used in other aMD studies of protein systems in aqueous solution:[27,30–33]

$$AlphaP = 0.2 \left[\frac{kcal}{mol}\right] \times N_{atoms} \qquad (1)$$

$$AlphaD = \frac{1}{5} \times 4 \left[\frac{kcal}{mol\ res}\right] \times N_{res} \qquad (2)$$

where in our case each BTA monomer has been considered as counting for 4 residues (one for the central ring + three side chains, see Figure 1) in the calculation of the extra bias for dihedral angles. This assumption has been made in light of the higher complexity and larger size of the water-soluble BTA monomers studied herein compared to standard amino acids composing proteins, on which aMD has been usually tested. In fact, initial tests counting each BTA monomer in the fiber as one single monomer produced negligible additional acceleration due to the extra bias applied on the dihedrals. The threshold values for the whole potential energy (*EthreshP*) and for the dihedral term (*EthreshD*) of the simulated systems have been set to:

$$EthreshP = Epot + AlphaP \qquad (3)$$

$$EthreshD = Edih + 5 \times AlphaD \qquad (4)$$

where *Epot* and *Edih* are the reference equilibrated values taken from our previous unbiased classical **MD₀** simulation.

In the MetaD simulations we have used as a single global variable the maximum *z* distance between the centers of mass of the BTA cores, calculated taking into account PBC, as defined in PLUMED2 as:

$$MAXZ = \beta \log \sum_i \exp(\frac{z_i}{\beta}) \qquad (5)$$

where $\beta$ has been set to 0.1 and the index *i* runs over all the BTA cores (48). We have set the hills height to 0.5 kcal/mol and the sigma to 0.05 nm. Hills were deposited every 500 simulation time steps (*i.e.*, every picosecond).

The average radius of the fiber has been estimated the following way. We have calculated the *z* component of the gyration tensor in the final state of the fiber, from which it is possible to derive an underestimate of the radius (approximating the shape of the fiber to a gyration cylinder). We have also calculated an upper limit of the fiber radius looking at the maximum coordinate values of the fiber in the *x* and *y* direction. We have taken the mean of these two extreme values for the estimation of our average radius, which has been compared to the experimental one (see Figure 3a).

The energetic analysis on the BTA fiber – *i.e.*, the electrostatic term of monomer-monomer interaction, $\Delta E_{ele}$ – has been conducted according to the MM-PBSA[53,54] approach on the last 100 ns of unbiased $MD_{post}$ conducted at the end of each simulation as:

$$\Delta E_{ele} = E_{ele}(fiber) - E_{ele}(monomer) \qquad (6)$$

where $E_{ele}$(fiber) is the average electrostatic energy of a BTA monomer into the fiber and $E_{ele}$(monomer) is the electrostatic energy of a monodisperse BTA monomer. Negative values for $\Delta E_{ele}$ thus indicate favorable electrostatic contribution for the BTAs to stay assembled in the fiber rather than disassembled in solution (see also SI). The variation between the $\Delta E_{ele}$ energy as measured during the 100 ns of unbiased $MD_{post}$ at the end of each *i* accelerated simulation and on the last 100 ns of classical **MD₀** was calculated as:

$$\Delta\Delta E_{ele} = \Delta E_{ele}(MD_{post})_i - \Delta E_{ele}(\mathbf{MD_0}) \qquad (6)$$

Negative values of $\Delta\Delta E_{ele}$ identify a better electrostatic state for the assembly reached during the accelerated simulations compared to the classical **MD₀** (see also SI).

The H-bonding analysis has been conducted using the *ptraj* module of AMBER 14, while SASA have been extracted using the *gmx_sasa* utility of Gromacs 5.1.2.

ACKNOWLEDGMENTS

The authors acknowledge the support from the Swiss National Science Foundation (SNSF grant: 2021_162827 to G.M.P.).

# REFERENCES

1. Yang, L., Tan, X., Wang, Z. & Zhang, X. Supramolecular Polymers: Historical Development, Preparation, Characterization, and Functions. *Chem. Rev.* **115,** 7196–7239 (2015).

2. De Greef, T. F. A. & Meijer, E. W. Materials science: supramolecular polymers. *Nature* **453,** 171–3 (2008).

3. De Greef, T. F. A. *et al.* Supramolecular polymerization. *Chem. Rev.* **109,** 5687–5754 (2009).

4. Aida, T., Meijer, E. W. & Stupp, S. I. Functional supramolecular polymers. *Science* **335,** 813–817 (2012).

5. Krieg, E., Bastings, M. M. C., Besenius, P. & Rybtchinski, B. Supramolecular Polymers in Aqueous Media. *Chem. Rev.* **16,** 2414–2477 (2016).

6. Torchi, A., Bochicchio, D. & Pavan, G. M. How the Dynamics of a Supramolecular Polymer Determines Its Dynamic Adaptivity and Stimuli-Responsiveness: Structure–Dynamics–Property Relationships From Coarse-Grained Simulations. *J. Phys. Chem. B* **122,** 4169–4178 (2018).

7. Lee, O. S., Stupp, S. I. & Schatz, G. C. Atomistic molecular dynamics simulations of peptide amphiphile self-assembly into cylindrical nanofibers. *J. Am. Chem. Soc.* **133,** 3677–3683 (2011).

8. Baker, M. B. *et al.* Consequences of chirality on the dynamics of a water-soluble supramolecular polymer. *Nat. Commun.* **6,** 6234 (2015).

9. Filot, I. A. W. *et al.* Understanding cooperativity in hydrogen-bond-induced supramolecular polymerization: A density functional theory study. *J. Phys. Chem. B* **114,** 13667–13674 (2010).

10. Kulkarni, C., Reddy, S. K., George, S. J. & Balasubramanian, S. Cooperativity in the stacking of benzene-1,3,5-tricarboxamide: The role of dispersion. *Chem. Phys. Lett.* **515,** 226–230 (2011).

11. Bejagam, K. K., Fiorin, G., Klein, M. L. & Balasubramanian, S. Supramolecular polymerization of benzene-1,3,5-tricarboxamide: A molecular dynamics simulation study. *J. Phys. Chem. B* **118,** 5218–5228 (2014).

12. Bejagam, K. K., Kulkarni, C., George, S. J. & Balasubramanian, S. External electric field reverses helical handedness of a supramolecular columnar stack. *Chem. Commun.* **51,** 16049–16052 (2015).

13. Lee, O. S., Cho, V. & Schatz, G. C. Modeling the self-assembly of peptide amphiphiles into fibers using coarse-grained molecular dynamics. *Nano Lett.* **12,** 4907–4913 (2012).

14. Garzoni, M., Cheval, N., Fahmi, A., Danani, A. & Pavan, G. M. Ion-selective controlled assembly of dendrimer-based functional nanofibers and their ionic-competitive disassembly. *J. Am. Chem. Soc.* **134,** 3349–3357 (2012).

15. Astachov, V. *et al.* In situ functionalization of self-assembled dendrimer nanofibers with cadmium sulfide quantum dots through simple ionic-substitution. *New J. Chem.* **40,** 6325–6331 (2016).

16. Bochicchio, D. & Pavan, G. M. From Cooperative Self-Assembly to Water-Soluble Supramolecular Polymers Using Coarse-Grained Simulations. *ACS Nano* **11,** 1000–1011 (2017).

17. Bochicchio, D. & Pavan, G. M. Effect of Concentration on the Supramolecular Polymerization Mechanism via Implicit-Solvent Coarse-Grained Simulations of Water-Soluble 1,3,5-Benzenetricarboxamide. *J. Phys. Chem. Lett.* **8,** 3813–3819 (2017).

18. Bochicchio, D., Salvalaglio, M. & Pavan, G. M. Into the dynamics of a supramolecular polymer at submolecular resolution. *Nat. Commun.* **8,** 147 (2017).

19. Albuquerque, R. Q., Timme, A., Kress, R., Senker, J. & Schmidt, H. W. Theoretical investigation of macrodipoles in supramolecular columnar stackings. *Chem. - A Eur. J.* **19,** 1647–1657 (2013).

20. Bejagam, K. K. & Balasubramanian, S. Supramolecular Polymerization: A Coarse Grained Molecular Dynamics Study. *J. Phys. Chem. B* **119,** 5738–5746 (2015).

21. Garzoni, M. *et al.* Effect of H-Bonding on Order Amplification in the Growth of a Supramolecular Polymer in Water. *J. Am. Chem. Soc.* **138,** 13985–13995 (2016).

22. Bochicchio, D. & Pavan, G. M. Molecular modelling of supramolecular polymers. *Adv. Phys. X*

**3,** 1436408 (2018).

23. Cantekin, S., de Greef, T. F. A. & Palmans, A. R. A. R. A. Benzene-1,3,5-tricarboxamide: a versatile ordering moiety for supramolecular chemistry. *Chem. Soc. Rev.* **41,** 6125–37 (2012).

24. Leenders, C. M. A. *et al.* Supramolecular polymerization in water harnessing both hydrophobic effects and hydrogen bond formation. *Chem. Commun.* **49,** 1963–1965 (2013).

25. Albertazzi, L. *et al.* Probing Exchange Pathways in One-Dimensional Aggregates with Super-Resolution Microscopy. *Science* **344,** 491–495 (2014).

26. Beltrán, E. *et al.* Self-organization of star-shaped columnar liquid crystals with a coaxial nanophase segregation revealed by a combined experimental and simulation approach. *Chem. Commun.* **51,** 1811–1814 (2015).

27. Hamelberg, D., Mongan, J. & McCammon, J. A. Accelerated molecular dynamics: A promising and efficient simulation method for biomolecules. *J. Chem. Phys.* **120,** 11919–11929 (2004).

28. Laio, A. & Parrinello, M. Escaping free-energy minima. *Proc. Natl. Acad. Sci. U. S. A.* **99,** 12562–6 (2002).

29. Barducci, A., Bonomi, M. & Parrinello, M. Metadynamics. *Wiley Interdiscip. Rev. Comput. Mol. Sci.* **1,** 826–843 (2011).

30. Pierce, L. C. T., Salomon-Ferrer, R., Augusto F. De Oliveira, C., McCammon, J. A. & Walker, R. C. Routine access to millisecond time scale events with accelerated molecular dynamics. *J. Chem. Theory Comput.* **8,** 2997–3002 (2012).

31. Miao, Y., Feixas, F., Eun, C. & McCammon, J. A. Accelerated molecular dynamics simulations of protein folding. *J. Comput. Chem.* **36,** 1536–1549 (2015).

32. Bucher, D., Grant, B. J., Markwick, P. R. & McCammon, J. A. Accessing a hidden conformation of the maltose binding protein using accelerated molecular dynamics. *PLoS Comput. Biol.* **7,** e1002034 (2011).

33. de Oliveira, C. A. F., Grant, B. J., Zhou, M. & McCammon, J. A. Large-scale conformational changes of Trypanosoma cruzi proline racemase predicted by accelerated molecular dynamics simulation. *PLoS Comput. Biol.* **7,** e1002178 (2011).

34. Hayder, M. *et al.* Three-dimensional directionality is a pivotal structural feature for the bioactivity of azabisphosphonate-capped Poly(PhosphorHydrazone) nanodrug dendrimers. *Biomacromolecules* **19,** 712–720 (2018).

35. Pavan, G. M., Barducci, A., Albertazzi, L. & Parrinello, M. Combining metadynamics simulation and experiments to characterize dendrimers in solution. *Soft Matter* **9,** 2593 (2013).

36. Hameau, A. *et al.* Theoretical and experimental characterization of amino-PEG-phosphonate-terminated polyphosphorhydrazone dendrimers: Influence of size and PEG capping on cytotoxicity profiles. *J. Polym. Sci. Part A Polym. Chem.* **53,** 761–774 (2015).

37. Jorgensen, W. L., Chandrasekhar, J., Madura, J. D., Impey, R. W. & Klein, M. L. Comparison of simple potential functions for simulating liquid water. *J. Chem. Phys.* **79,** 926 (1983).

38. Case, D.A. *et al.* AMBER 14, University of California, San Francisco (2014).

39. Miao, Y. *et al.* Improved reweighting of accelerated molecular dynamics simulations for free energy calculation. *J. Chem. Theory Comput.* **10,** 2677–2689 (2014).

40. Jing, Z. & Sun, H. A Comment on the Reweighting Method for Accelerated Molecular Dynamics Simulations. *J. Chem. Theory Comput.* **11,** 2395–2397 (2015).

41. Amado Torres, D., Garzoni, M., Subrahmanyam, A. V., Pavan, G. M. & Thayumanavan, S. Protein-triggered supramolecular disassembly: Insights based on variations in ligand location in amphiphilic dendrons. *J. Am. Chem. Soc.* **136,** 5385–5399 (2014).

42. Munkhbat, O., Garzoni, M., Raghupathi, K. R., Pavan, G. M. & Thayumanavan, S. Role of Aromatic Interactions in Temperature-Sensitive Amphiphilic Supramolecular Assemblies. *Langmuir* **32,** 2874–2881 (2016).

43. Papmeyer, M., Vuilleumier, C. A., Pavan, G. M., Zhurov, K. O. & Severin, K. Molecularly Defined Nanostructures Based on a Novel AAA-DDD Triple Hydrogen-Bonding Motif. *Angew. Chemie - Int. Ed.* **55,** 1685–1689 (2016).

44. Berendsen, H. J. C., van der Spoel, D. & van Drunen, R. GROMACS: A message-passing parallel molecular dynamics implementation. *Comput. Phys. Commun.* **91,** 43–56 (1995).

45. Abraham, M. J. *et al.* Gromacs: High performance molecular simulations through multi-level parallelism from laptops to supercomputers. *SoftwareX* **1–2,** 19–25 (2015).

46. Tribello, G. A., Bonomi, M., Branduardi, D., Camilloni, C. & Bussi, G. PLUMED 2: New feathers for an old bird. *Comput. Phys. Commun.* **185,** 604–613 (2014).

47. Berendsen, H. J. C., Postma, J. P. M., Van Gunsteren, W. F., Dinola, A. & Haak, J. R. Molecular dynamics with coupling to an external bath. *J. Chem. Phys.* **81,** 3684–3690 (1984).

48. Grest, G. S. & Kremer, K. Molecular dynamics simulation for polymers in the presence of a heat bath. *Phys. Rev. A* **33,** 3628–3631 (1986).

49. Bussi, G., Donadio, D. & Parrinello, M. Canonical sampling through velocity rescaling. *J. Chem. Phys.* **126,** 14101 (2007).

50. Kräutler, V., Van Gunsteren, W. F. & Hünenberger, P. H. A fast SHAKE algorithm to solve distance constraint equations for small molecules in molecular dynamics simulations. *J. Comput. Chem.* **22,** 501–508 (2001).

51. Hess, B., Bekker, H., Berendsen, H. J. C. & Fraaije, J. G. E. M. LINCS: A Linear Constraint Solver for molecular simulations. *J. Comput. Chem.* **18,** 1463–1472 (1997).

52. Darden, T., York, D. & Pedersen, L. Particle mesh Ewald: An N·log(N) method for Ewald sums in large systems. *J. Chem. Phys.* **98,** 10089–10092 (1993).

53. Kollman, P. A. *et al.* Calculating structures and free energies of complex molecules: Combining molecular mechanics and continuum models. *Acc. Chem. Res.* **33,** 889–897 (2000).

54. Srinivasan, J., Cheatham, T. E., Cieplak, P., Kollman, P. A. & Case, D. A. Continuum solvent studies of the stability of DNA, RNA, and phosphoramidate - DNA helices. *J. Am. Chem. Soc.* **120,** 9401–9409 (1998).

Supporting information for:

# Accelerated Atomistic Simulations of a Supramolecular Polymer in Water

*Davide Bochicchio and Giovanni M. Pavan\**

Department of Innovative Technologies, University of Applied Sciences and Arts of Southern Switzerland, Galleria 2, CH-6928 Manno, Switzerland

giovanni.pavan@supsi.ch

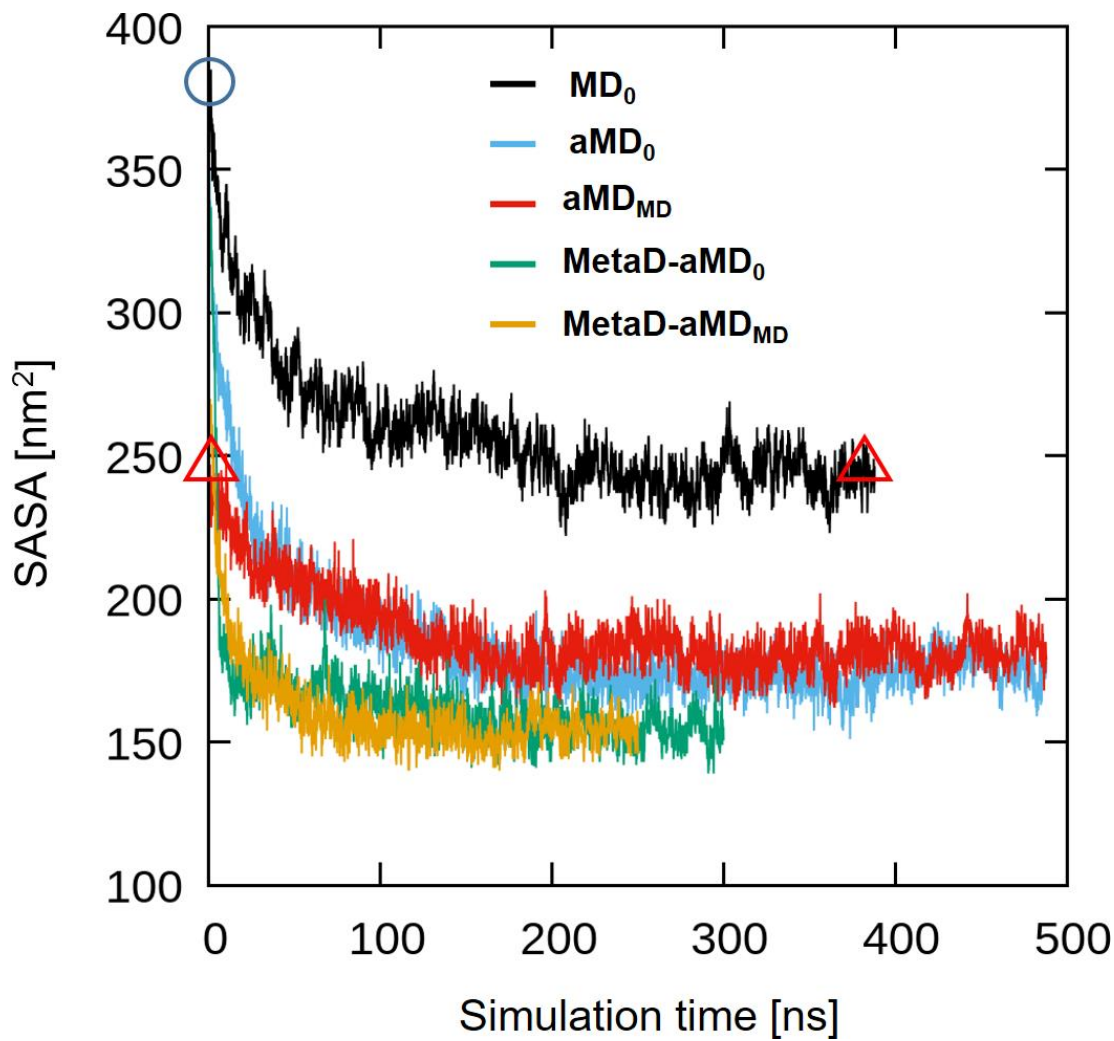

**Figure S1**. Solvent accessible surface area (SASA) of the BTA fibers as a function of the simulation time. Reference **MD$_0$** data is shown in black. Among accelerated simulations, **aMD$_0$** (cyan) and **MetaD-aMD$_0$** (green) started from the same extended initial configuration for the fiber used in **MD$_0$** (blue circle), while **aMD$_{MD}$** (red) and **MetaD-aMD$_{MD}$** (yellow) started from the pre-equilibrated folded configuration of the fiber obtained at the end of **MD$_0$** (red triangle). The last 100 ns of each data pertain to the unbiased classical MD$_{post}$ run following each biased simulation.

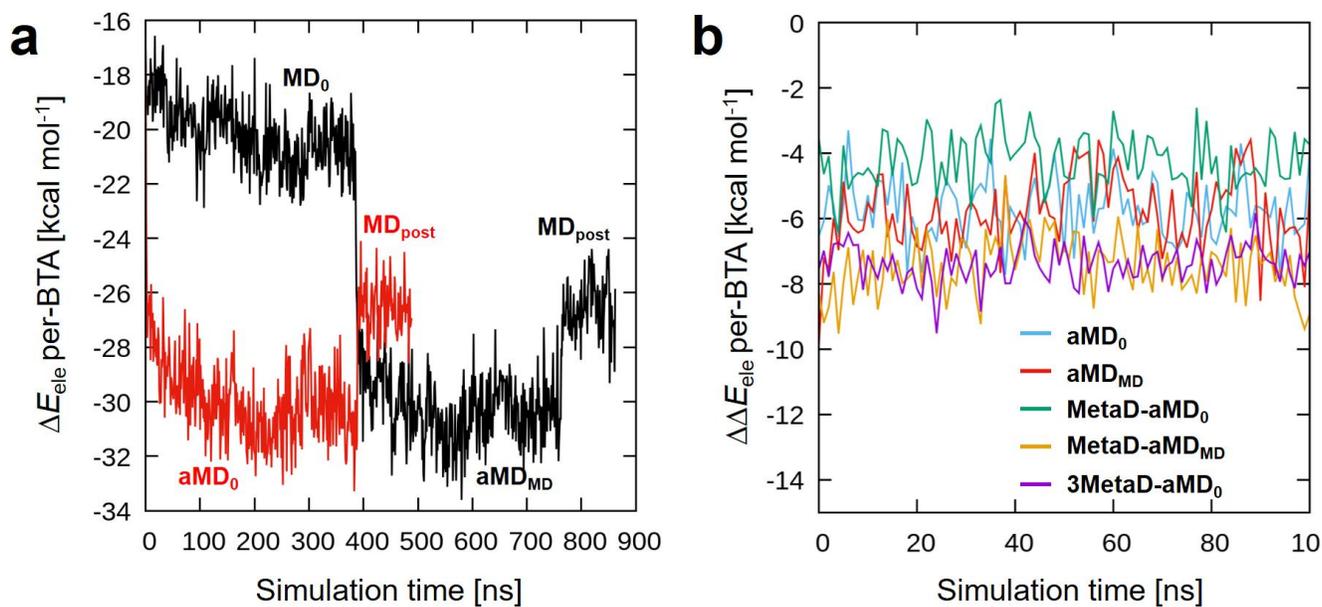

**Figure S2.** Evolution of electrostatic self-assembly energy in the fiber. (a) Electrostatic component of BTA self-assembly energy ($\Delta E_{ele}$) as a function of simulation time – black curve: **MD$_0$** (0-400 ns), followed by **aMD$_0$** and the last 100 ns of unbiased MD$_{post}$. Red curve: **aMD$_0$** followed by the last 100 ns of unbiased MD$_{post}$. The two red and black MD$_{post}$ $\Delta E_{ele}$ energies are nearly identical and more favorable than in **MD$_0$**. (b) Improvement of the electrostatic self-assembly energy ($\Delta\Delta E_{ele}$) between the MD$_{post}$ phase following each biased simulation and the last 100 ns of **MD$_0$**: $\Delta\Delta E_{ele} = \Delta E_{ele}(MD_{post}) - \Delta E_{ele}(MD_0)$. All values are expressed per-BTA monomer.

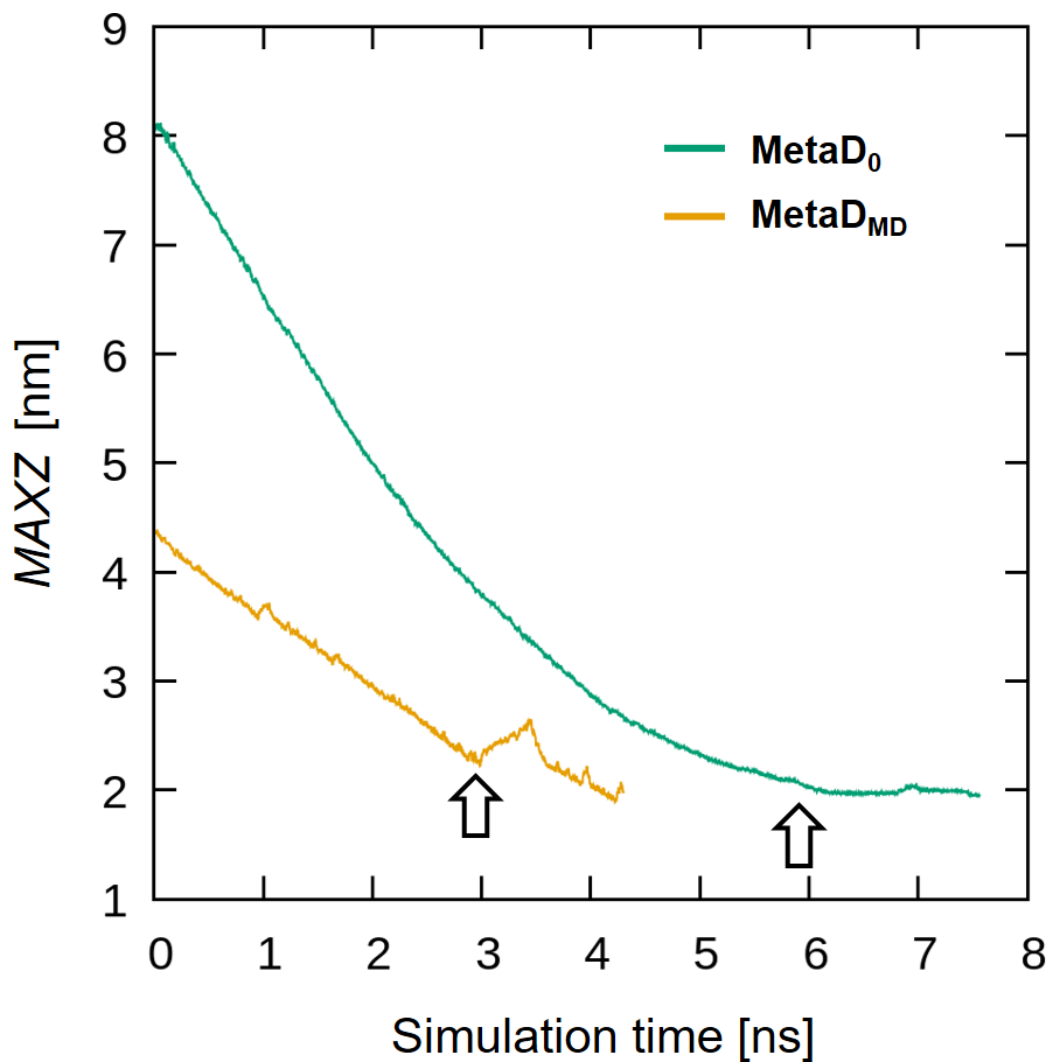

**Figure S3.** Evolution of the collective variable (*MAXZ*) used to accelerate fiber folding during the MetaD runs (**MetaD$_0$**: green; **MetaD$_{MD}$**: yellow). The arrows indicate the points along the MetaD simulations where we extracted the pre-folded configurations for the BTA fiber that have been then relaxed by means of aMD.